\documentstyle[11pt,newpasp,twoside,epsf,psfig]{article}
\def\edcomment#1{\iffalse\marginpar{\raggedright\sl#1\/}\else\relax\fi}
\reversemarginpar

\begin{document}
\title{Bragg gratings in multi-mode fiber optics for
wavelength calibration of GAIA and RAVE spectra}
 \author{Claudio Pernechele and Ulisse Munari}
\affil{INAF - Astronomical Observatory of Padova, vicolo Osservatorio 5,
I-35122 - Padova - Italy}

\begin{abstract}
We propose a new technique, the use of FBGs (fiber Bragg gratings), for
accurate, easy and low cost wavelength calibration of GAIA, RAVE and
follow-ups spectra at local Observatories. FBGs mark the spectra with
absorption lines, freely defined in number and position during the fibers
manufacturing. The process goes in parallel with the science exposure and
through the same optical train and path, thus ensuring the maximum return in
wavelength calibration accuracy. Plans to manufacture and test FBGs for the
CaII/Paschen region are underway at the Astronomical Observatory of Padova.
\end{abstract}

\section{Introduction}

Modern and demanding application of radial velocities, like those associated
with the search for extra-solar planets, have promoted the introduction of
new methods of wavelength calibration.

The key to success has been recording of the reference wavelength grid
for to the whole duration of the science exposure and through the 
same optical train and path. Two approaches have been followed: the {\em
parallel} (where Thorium lines going through telescope optics are recoded on
the CCD close and parallel to the science spectrum with an exposure lasting
as long as the science observation), and the {\em Iodine-cell} one (the
stellar spectrum passes through and is absorbed by a cell filled with Iodine
vapors, that mark the stellar spectrum with a huge number of sharp and
equally intense absorption lines). Both approaches require delicate
instrumentation, careful maintenance and dedicated spectrographs. 
As such they are not easily manageable on small, all-purpose telescopes.

We propose here a third, independent way to mark with reference wavelengths
the science spectrum as it is recorded, through exactly the same optical
train and path. It consists of using optical fibers with recorded in them
Bragg gratings that allows undisturbed passage of all wavelengths except a
selection of few ones that will appear in the spectrum as sharp, deep
absorption lines of constant and pre-selected wavelengths. The latter are
then used as a reference grid to accurately wavelength calibrate the science
spectrum.

The system is easy and fast, allows to decouple the spectrograph from the
telescope and operate it in a controlled environment, requires minimal
mechanical work to be adapted to existing spectrographs, the marking
absorption lines can be placed at any preferred wavelength, and the fibers
can be mass-produced at low cost for distribution to interested
Observatories. FBGs can be easily installed on spectrographs already
designed to operate in multiple fiber mode. The RAVE project (Steinmetz 2003),
with its fixed wavelength range, would represent an ideal case for the
application of FBGs.

The FBGs (fiber Bragg grating) appear of potential interest also to GAIA. 
The latter's spectrograph operates in TDI, slit-less mode with thousands of
spectra simultaneously crossing the field of view as the satellite spins. No
conventional calibration lamp can be used, obviously. The base-lined
wavelength calibration procedure foresees using stars of recognized constant
radial velocity as they travel across the spectrograph focal plane, by
accurately linking their astrometric position on the sky to the wavelength
of their absorption lines on the pixels of the CCDs. 

A bundle of FBGs properly introduced in the GAIA spectrograph optical train
would mark with a set of properly placed reference absorption lines the
spectra of all stars entering the field of view. The FBGs could become the
prime wavelength calibration mean of GAIA spectra (particularly for
preliminary solutions while the mission is still flying), or a backup for
the method based on standard radial velocity stars in case this would not
satisfactorily operate for whatever reason once the satellite will reach
L$_2$. The FBGs dissipate no power and have no movable parts. They should
also stand the low operating temperatures of the satellite in L$_2$ and
should do not degrade with time, and thus looking ideal for use in space.

No FBG has ever been used as proposed in this paper, and thus confirmatory
tests are required. At the Astronomical Observatory of Padova we plan to
manufacture and test prototype fibers for the GAIA/RAVE wavelength range to
check their performance with actual observations at the telescope.

\section{Fiber Bragg Grating principle}

Fiber gratings are made by laterally exposing the core of the fiber to a
periodic pattern of intense ultraviolet light. The exposure produces a
permanent increase in the refractive index of the fiber's core, creating a
fixed index modulation according to the exposure pattern: this fixed pattern
is called a grating. At each periodic change of refraction, a small amount
of light is reflected (Bragg diffraction). All the reflected light signals
combine coherently to one large reflection at a particular wavelength when
the grating period is approximately half the input light wavelength. This
is referred as the Bragg condition and the wavelength at which this
reflection occurs is called the Bragg wavelength ($\lambda_B$). Light at
wavelengths different from $\lambda_B$, which are not phase matched, will
pass unabsorbed through the fiber. Thus, light propagates through the
grating with negligible attenuation and only those wavelengths that satisfy
the Bragg condition are affected and strongly back-reflected.

Application of this technology to astronomical spectra at GAIA wavelengths
is attractive for two basic reasons: ($a$) the technology is already mature
with the tele-communication industry that operate it in the near-IR, down to
1~$\mu$m, and ($b$) the limited $\bigtriangleup \lambda$ range observed by
GAIA and RAVE requires only a limited number of reference lines to be
properly calibrated (see discussion in Desidera and Munari 2003), which in turn
means an easier FBGs production.

\begin{figure}[!t]
\centerline{\psfig{file=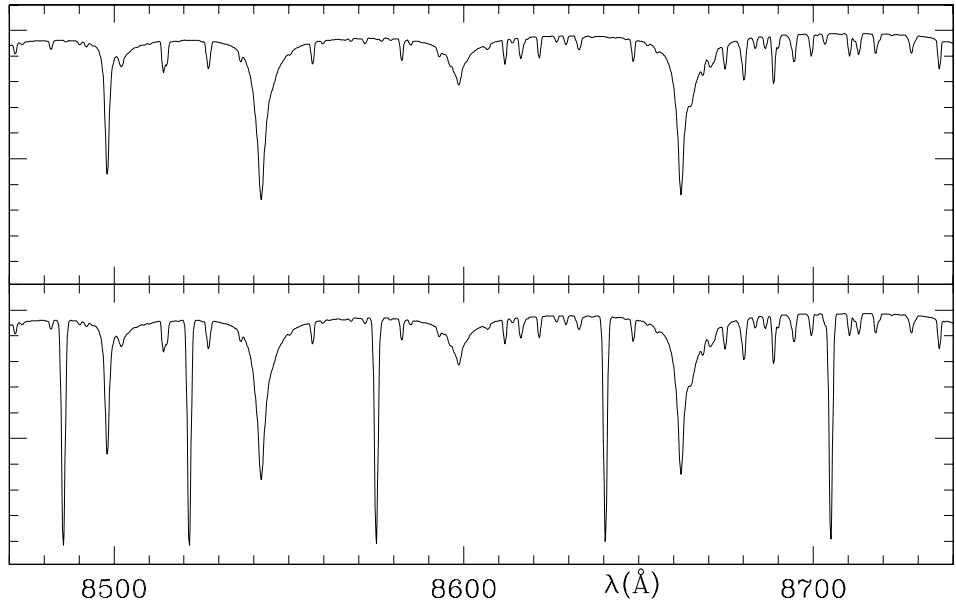,height=8cm}}
\caption{The spectrum of an F3~V star before and after passing through an
hypotetical FBG fiber marking on it 5 reference wavelengths.}
\end{figure}

\section{Fiber Bragg Grating production}

Exposing a photosensitive fiber, usually Germanium-doped (Yuen, 1982), to an
intensity pattern of ultraviolet radiation (at wavelengths between 240 and 255 nm,
where the Germanium has an absorption band) is used to produce (i.e. write) a
fiber Bragg grating. In its basic form the grating selectively reflects
light at the Bragg wavelength $\lambda_B = 2nH$, where $H$ is the grating
spacing and $n$ is the effective index of refraction. Both $H$ and $n$
depend on fabrication parameters: $n$ depend on the modal dispersion
characteristics of the fiber and lies, for each mode, between the core and
cladding indices. The functional form of the normalized
frequency ($\nu$) parameter is given by:
\begin{equation}
\nu = {{\pi d} \over {\lambda}} \sqrt{({n_{co}}^2-{n_{cl}}^2)}
\end{equation}
\noindent
where $d$ is the diameter of the fiber core, $\lambda$ is the free space wavelength, $n_{co}$ the core 
index and $n_{cl}$ the clad index. Note that the difference between the two squared index 
is also the numerical aperture of the fiber. For low enough $\nu$ values the fiber is 
single mode and only one Bragg reflection is observed at $\lambda_B $
wavelength given above. At a certain value of $\nu$ the fiber become two moded with two 
separated Bragg wavelengths, and for larger $\nu$ values many more modes begin to appear 
in the fiber grating spectrum. The lowest order or fundamental mode has the longest 
$\lambda_B$, and the higher order modes have progressively shorter $\lambda_B$ with the cladding index 
giving the limit on the shortest $\lambda_B$. The peak reflectivity is an increasing 
function of the grating length and of the difference between $n_{co}$ and $n_{cl}$ 
(Lam \& Garside, 1981).

\begin{figure}[!t]
\plottwo{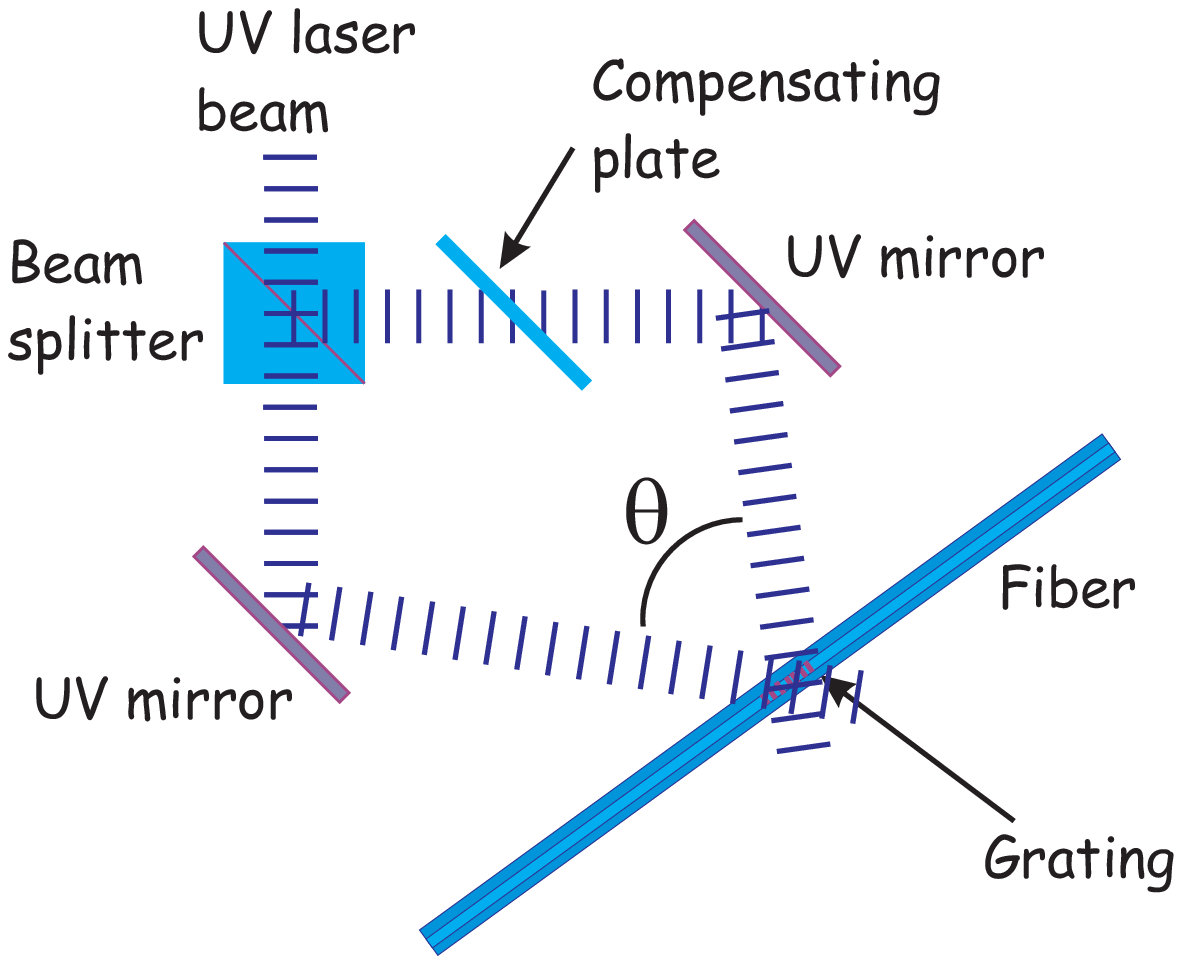}{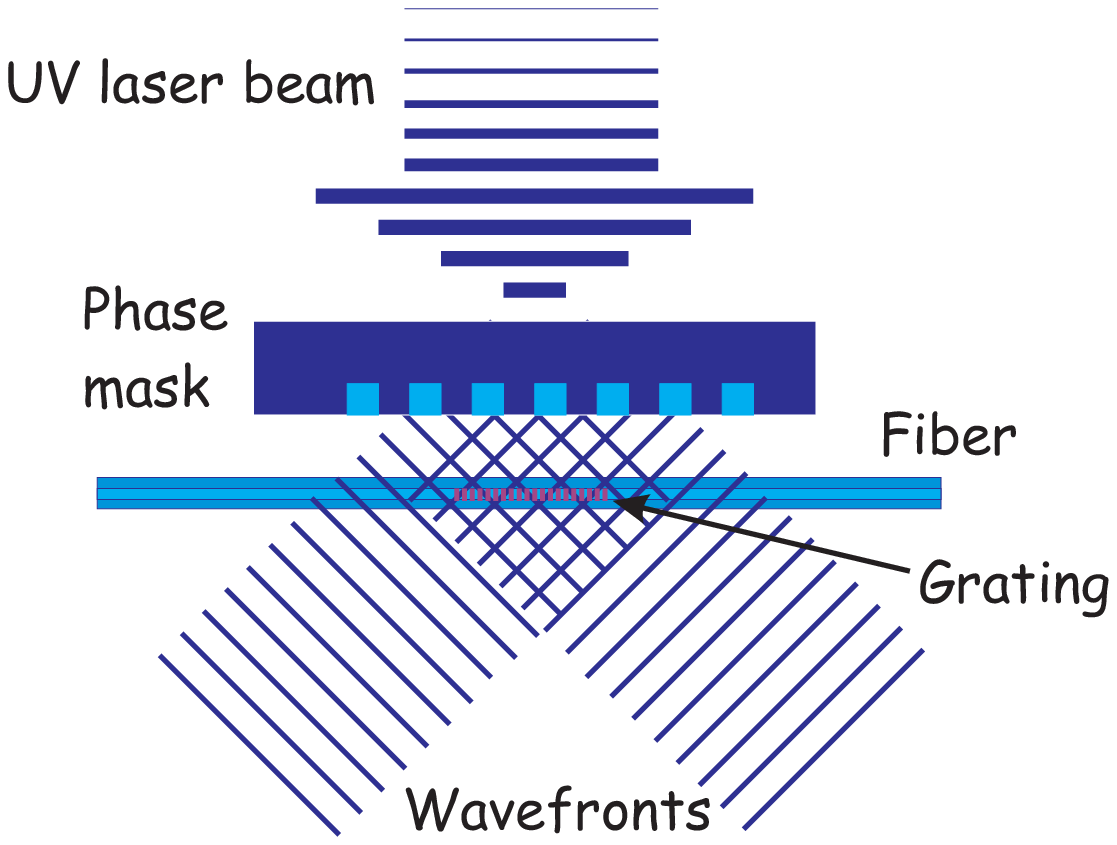}
\caption{Writing the FBG via bulk interferometry (left panel) or phase mask (right panel).}
\end{figure}

Recording FBGs inside a fiber usually involves interference between two coherent UV 
laser beams that precisely define the spacing between the grating planes. For 
symmetric incidence and an angle $\theta$ between the UV beams, the pitch of the recorded 
grating is 0.5$ \lambda sin(\theta/2)$, where $\lambda$ is the illumination wavelength. 

The basic interferometric setup splits the laser beam to create an
interference pattern on the fiber (Morey et al., 1989) as shown in Figure 2
(left panel). Because the beams travel different paths, the laser must have
sufficient temporal coherence to account for path-length mismatch, and good
spatial coherence. The sensitivity to mechanical instabilities of this
method of FBG production can make difficult to produce identical replicas.

The other way for writing FBGs is to use phase mask to generate multiple
beams (Hill et al., 1993). Two of the beams carry approximately 40\% of the
total energy and closely overlap the phase mask surface to create an
interference pattern at the desired Bragg wavelength. The fiber-phase mask
assembly can be illuminated either by a large beam to cover the full fiber
Bragg grating length or by a small scanning beam. Because the fiber is
usually in proximity to the phase mask, the assembly is a very stable
mechanical system, suitable for mass-production of identical FBG replicas. 
The pattern recorded into the fiber is a copy of the phase mask (pitch and
chirp) scaled by about 50\%.

Typically two types of UV laser are used to manufacture FBGs: continuous-wave 
frequency-doubled argon-ion lasers at 244 nm, and pulsed excimer lasers operating 
at either 248 nm (KrF) or 193 nm (ArF).

\end{document}